\documentclass{sig-alternate}
\usepackage[hang,flushmargin,stable]{footmisc}
\usepackage{graphicx}
\usepackage{amssymb}
\usepackage{amsmath}
\usepackage{colortbl}
\usepackage{color}
\usepackage{indentfirst}
\usepackage{url}
\usepackage{subfigure}
\usepackage{multirow}
\usepackage{threeparttable}
\usepackage{tikz}
\usepackage{bbding}
\usepackage[all]{xy}
\usepackage{algorithmic}
\usepackage{array}
\usepackage{multirow}
\usepackage{flushend}
\usepackage{paralist}
\usepackage{tweaklist}
\usepackage[dvips, bookmarks=false, colorlinks=true, plainpages = false,
  linkcolor = blue,   citecolor = blue, urlcolor = blue, filecolor = blue]{hyperref}

\renewcommand{\itemhook}{\setlength{\topsep}{2pt}%
  \setlength{\itemsep}{0pt}}
\renewcommand{\enumhook}{\setlength{\topsep}{2pt}%
  \setlength{\itemsep}{0pt}}

\renewcommand{\scriptsize}{\fontsize{7.5}{9.5}\selectfont}

\makeatletter
\let\@copyrightspace\relax
\makeatother

\renewenvironment{thebibliography}[1]{%
\begin{oldthebibliography}{#1}%
\setlength{\parskip}{0ex}%
\setlength{\itemsep}{0ex}%
}{\end{oldthebibliography}}

\newcommand{\naive}{na\"\i ve}
\newcommand{\descr}[1]{\vspace{0.18cm} \noindent \textbf{#1}}

\newcommand{\spa}{\mbox{ }}
\newcommand{\equal}{\hspace*{-0.05cm}=\hspace*{-0.05cm}}
\newcommand{\Cset}{\ensuremath{\mathcal{C}}}
\newcommand{\Sset}{\ensuremath{\mathcal{S}}}
\newcommand{\Z}{\mathbb{Z}}
\newcommand{\Client}{{\sf  Client}}
\newcommand{\Server}{{\sf  Server}}
\newcommand{\CA}{{\sf \small CA}}

\newcommand{\FSG}{fully sequenced genome}
\newcommand{\FSGS}{fully sequenced genomes}
\newcommand{\FGS}{full genome sequencing}

\urldef\gattaca\url{http://www.imdb.com/title/tt0119177/}

\begin{document}
\pagenumbering{arabic}
\thispagestyle{plain}



\title{Countering GATTACA\hspace{-0.07cm}{\huge \bf \thanks{$\;\;$See \gattaca.}}\hspace{0.05cm} :
Efficient and Secure Testing of Fully-Sequenced Human
Genomes (Full Version){\huge \bf \thanks{$\;\;$A preliminary version of this paper appeared in \href{http://www.sigsac.org/ccs/CCS2011/}{\textsf{ACM CCS 2011.}} The present version supercedes 
it. New material includes new techniques for privacy-preserving paternity testing in Section~\ref{sec:new}.}}}

\author{Pierre Baldi{\large $^{\dag,\ddag}$} \hspace{0.12cm} Roberta Baronio{\large $^\dag$} \hspace{0.12cm}
Emiliano De Cristofaro{\large $^\ddag$} \hspace{0.12cm} Paolo Gasti{\large $^\ddag$} \hspace{0.12cm} Gene Tsudik{\large $^\ddag$}\vspace{0.15cm}\\
{\large $^\dag$}  \fontfamily{phv}\selectfont Institute for Genomics and Bioinformatics \hspace{0.3cm} {\large $^\ddag$}  \fontfamily{phv}\selectfont Department of Computer Science \vspace{0.1cm}\\
\email{{\fontfamily{phv}\selectfont{\{pfbaldi,rbaronio,edecrist,pgasti,gts\}@uci.edu}}} \vspace{0.1cm}\\
{\fontfamily{phv}\selectfont University of California, Irvine}}

\maketitle

\begin{abstract}
Recent advances in DNA sequencing technologies have put ubiquitous availability of fully sequenced 
human genomes within reach. It is no longer hard to imagine the day when everyone will have the means to 
obtain and store one's own DNA sequence. Widespread and affordable availability of \FSGS\ immediately opens up 
important opportunities in a number of health-related fields. In particular, common genomic applications and tests 
performed {\em in vitro} today will soon be conducted computationally, using digitized genomes. New applications will be 
developed as genome-enabled medicine becomes increasingly preventive and personalized. However, this progress 
also prompts significant privacy challenges associated with potential loss, theft, or misuse of genomic data.
In this paper, we begin to address genomic privacy by focusing on three important applications: {\em Paternity Tests}, 
{\em Personalized Medicine}, and {\em Genetic Compatibility Tests}. After carefully analyzing these applications and 
their privacy requirements, we propose a set of efficient techniques based on  {\em private set operations}. 
This allows us to implement in {\em in silico}  some operations that are currently performed via {\em in vitro} methods,
in a secure fashion. Experimental results demonstrate that proposed techniques are both feasible and practical today.
\end{abstract}

%
%

\section{Introduction}\label{sec:intro}
Over the past four decades, DNA sequencing has been one of the major driving forces in life-sciences, producing
full genome sequences of thousands of viruses and bacteria, and dozens of eukaryotic organisms, from yeast to
man (e.g., \cite{goffeau1996lg, adams2000gsd, waterston2002isa, humangenome01}). This trend is only being
accentuated by modern High-Throughput Sequencing (HTS) technologies: the first diploid human genome
sequences were recently produced \cite{levy07, wheeler08, Wang_2008} and a project to sequence 1,000 human
genomes has been essentially completed \cite{kaiser08, new1, new2}.  Different HTS technologies are competing
to sequence an individual human genome --- composed of about 3 billion DNA nucleotides (or bases) --- for less than 
\$1,000 by 2012 \cite{service06}, and even less than \$100 five years later, reaching the point  where human genome
sequencing will be a commodity costing less than an X-ray or an MRI scan.
Ubiquity of human and other genomes creates enormous opportunities and challenges. In particular, it
promises to address one of the greatest societal challenges  of our time: the unsustainable rise of health care costs, by ushering a 
new era of genome-enabled predictive, preventive, participatory,  and personalized medicine (``P4'' medicine). In time, 
genomes could become part of the {\em Electronic Medical Record}  of every individual \cite{hoffman2007genome}.

However, widespread availability of HTS technologies and genomic data 
exacerbates ethical, security, and privacy concerns \cite{ collins2001implications}.
A full genome sequence not only uniquely identifies each one of us; it also contains information about, for instance,
our ethnic heritage, disease predispositions, and many other phenotypic traits~\cite{correlated,scientificmatch}. 
Traditional approaches to privacy, such as de-identification or aggregation
\cite{malin2005,homer2008}, become completely moot in the genomic era, since the genome itself is the ultimate identifier.
To further compound the privacy problem, health information is increasingly shared electronically among insurance
companies, health care providers and employers. This, coupled with the possibility of
creating large centralized genome repositories, raises the specter of possible abuses.

Some federal laws have been passed to begin addressing privacy issues. The 2003 Health Insurance Portability and
Accountability Act (HIPAA) provides a general framework for protecting and sharing Protected Health Information (PHI)
\cite{durham2002note, jennifer2003new, mcguire2006column}. In 2008, the Genetic Information
Nondiscrimination Act (GINA) 
was adopted to prohibit discrimination on the basis of genetic
information. with respect to health insurance and employment \cite{wadman08}. While providing general guidelines and
a basic safety net, current legislation does not offer detailed technical information about safe and 
privacy-preserving ways for storing and querying genomes. In short, technical issues of security and privacy for 
HTS and genomic data remain both important and relatively poorly understood.

While privacy issues are not yet hampering progress in basic genomic research, it is not too early to start
investigating them, particularly, in light of their complexity, potential impact on society, and current efforts to
reform the health care system. It remains unclear where personal genomic information will be stored, who 
will have access to it, and how it will be queried and shared. To remain flexible, we can imagine a general 
framework comprised of two kinds of basic entities:
(1) Data Centers where genomic data is stored, and (2) Agents/Agencies interested in querying this data.
Granularity of Data Centers could vary. At one end of the spectrum, every individual could be her
own Data Center and store the genome on a personal computer, cell phone, or some other device. At the other
extreme, we could envision national or even international Data Centers storing millions (or even billions) of genomic
sequences. Data Centers could also be envisioned at the granularity of family, school, pharmacy, laboratory, hospital,
city, county or state. Likewise, many different types of Agents/Agencies are conceivable, ranging from individuals and
personal physicians, to family members, pharmacies, hospitals, insurance companies, employers and 
government agencies (e.g., the FBI), or international organizations. Various Agents/Agencies might be allowed to 
query different aspects of genomic data and might be required to satisfy different query privacy requirements. 
In addition, one could imagine cases (e.g., criminal search or proprietary diagnostic technology) where both
the genomic data and queries against it must remain private.

The main security and privacy challenge is how to support such queries with low storage costs and reasonably short query times, while satisfying privacy and security requirements
associated with a given type of transaction. Unfortunately, current methods for privacy-preserving data 
querying do not scale to genomic data sizes. 
Several cryptographic techniques have been proposed that --- though not addressing the 
case of fully-sequenced genomes --- focus on private computation over genomic fragments. Specifically, they allow two or more parties 
to engage in protocols that reveal only the end-result of a given computation on their respective genomic data, 
without leaking any additional information. The main thrust of this paper is to adapt and deploy efficient cryptographic 
techniques 
to address specific genomic queries and applications, described below.

\subsection{Applications}\label{sec:applications}
As mentioned above, availability of affordable \FGS{} makes it increasingly possible to query and test 
genomic information not only {\em in vitro},  but also {\em in silico} using computational techniques.  
We consider three concrete examples of such tests and corresponding privacy-relevant scenarios.

\descr{Paternity Tests} establish whether a male individual is the biological father of another
individual, using genetic fingerprinting. Advances in biotechnology facilitated DNA paternity tests
and stimulated the creation of  hundreds of online companies offering testing via self-administered cheek swabs
for as little as~\$79 (e.g., \url{http://www.gtldna.net}). However, this practice raises several security and privacy concerns: 
the testing company must be trusted with privacy and accuracy of test results, as well as
with swabs that might yield \FGS{}. We believe that, ideally, any two individuals, in possession
of their genomes should be able to conduct  a privacy-preserving paternity test with no involvement
of any third parties. Only the outcome of the test ought to be learned by one or both parties
and no other sensitive genomic information should be disclosed. 

\descr{Personalized Medicine} is recognized as a significant \linebreak paradigm shift and a major trend in health care, 
moving us closer to a more precise, powerful, and holistic type of medicine~\cite{weston2004}.
With personalize medicine, treatment and medication type/dosage would be tailored to the precise
genetic makeup of individual patient.
For example, measurements of {\em erbB2} protein in breast, lung, or colorectal cancer patients are taken before
selecting proper treatment. It has been showed that the trastuzumab monoclonal antibody
is effective only in patients whose genetic receptor is over-expressed~\cite{trastuzumab}.
Furthermore, the FDA has recently recommended testing for the thiopurine S-methyltransferase ({\em tpmt})
gene, prior to prescribing for 6-mercaptopurine and azathioprine --- two drugs used for treating childhood \linebreak leukemia
and autoimmune diseases. The {\em tpmt} gene codes for the TPMT enzyme that metabolizes thiopurine drugs:
genetic polymorphisms affecting enzymatic activity are correlated with variations in sensitivity and toxicity
response to such drugs.
Patients suffering from this genetic disease (1 in 300) only need 6-10\% of the standard dose of thiopurine drugs;
if treated with the full dose, they risk severe bone marrow suppression and subsequent death~\cite{abbott2003}.
Not surprisingly, experts predict that availability of \FGS{} 
will further stimulate development of personalized medicine~\cite{ginsburg2009}.

\descr{Genetic Tests} are  routinely used for several purposes,
such as newborn screening, confirmational diagnostics, as well as 
pre-symp\-to\-mat\-ic testing, e.g., predicting Huntington's disease \cite{nature1983} and
estimating risks of various types of cancer. We focus on genetic {\em compatibility} tests, 
whereby potential or existing partners wish to assess the possibility of
transmitting to their children a genetic disease with Mendelian inheritance~\cite{mendelian}.
Modern genetic testing can accurately predict whether a couple is at risk of conceiving a
child with an autosomal recessive disease. Consider, for instance, {\em Beta-Thalassemia minor}, that causes red cells to 
be smaller than average, due to a mutation in the {\em hbb} gene. It is called {\em minor} when
the mutation occurs only in one allele. This {\em minor} form has no severe  impact on a subject's quality of life.
However, the {\em major} variant --- that occurs when both alleles carry the mutation --- is
likely to result in premature death, usually, before age twenty. Therefore, if both partners 
silently carry the {\em minor} form, there is a 25\% chance that their child could carry the major variety.
Another example is the {\em Lynch Syndrome} (also known as Hereditary Nonpolyposis Colon Cancer), a 
genetic condition --- most commonly inherited from a parent --- 
associated with the high risk of colon cancer~\cite{kastrinos2009}.
Parents with this syndrome have a 50\% chance of passing it on to their children.
Since the possibility of inheritance is maximized if both parents carry the mutations,
testing for Lynch Syndrome is crucial.

\descr{Note on Non-human Genomes:} Although this paper focuses on human genomes, some 
aforementioned scenarios apply to other organisms, e.g., crops and animals~\cite{beckmann1986}.
For instance, a paternity test may certify a purebred dog's bloodline or genetic tests
may determine the quality of a racing horse. In fact, DNA ``barcodes'' identifiers are already
embedded in genomes of genetically modified species. Conceivably, future veterinary treatments
may also involve elements of personalized medicine for animals.

\subsection{Roadmap}
Motivated by the emerging affordability of \FGS, we combine domain knowledge
in biology, genomics, bioinformatics, security, privacy and applied cryptography in order
to better understand the corresponding security and privacy challenges. In particular, we analyze
specific requirements of three types of applications discussed above:
Paternity Tests, Personalized Medicine and Genetic Tests. In the process, we carefully
consider today's {\em in vitro} procedure  for each application and
analyze its security and privacy requirements in the digital domain. This type of approach
allows us to gradually craft specialized protocols that incur
appreciably lower overhead than state-of-the-art. However, as is well known,
``lower overhead'' does not necessarily imply practicality. Therefore, we demonstrate ---
via experiments on commodity hardware --- that proposed protocols are indeed
viable and practical {\em today}. Source code of our implementations is publicly
available.
We hope that it can help in developing privacy-aware operations on full genomes and
allows individuals (in possession of their sequenced genomes) to run genetic tests with privacy. 

\descr{Organization.} We overview related work in the next section. Then, Sec.~\ref{prelim} introduces  
biological and cryptographic background used throughout
the rest of the paper. The core of the paper is in Sec.~\ref{sec:our} that includes step-by-step
design of protocols for each aforementioned application. It also presents experimental
results. Next, Sec.~\ref{sec:security} provides security arguments for proposed protocols, followed
by the summary and the discussion of future work in Sec.~\ref{conc}.

\section{Related Work}\label{sec:related}
As discussed in Section~\ref{sec:intro}, traditional approaches to privacy, such as de-identification,
are often ineffective on genomic data, since the genome itself is the ultimate identifier.
We refer to~\cite{malin2005,homer2008,wang2009learning,zhouESORICS} for details on privacy risks associated to 
releasing genomic information, even when aggregated.
Motivated by the sensitivity of genomic information, the security research community has begun to
develop mechanisms to enable secure computation on genomic data. A number of cryptographic
protocols have been proposed for private searching, matching and evaluating similarity of \linebreak strings,
including DNA sequences. Also, prior work has considered specific (privacy-preserving) genomic
operations. This section overviews relevant prior results and highlights their potential limitation.

\smallskip
\subsubsection*{Searching and Matching DNA}
Troncoso-Pastoriza, et al.~\cite{ccs07} proposed an error-resilient privacy-preserving protocol
for string searching.
In it, one party (e.g., Alice), with her own DNA snippet, can verify the existence of a
short template (e.g., a genetic test held by a service provider -- Bob) within her snippet.
This technique handles errors and maintains privacy of both the template and the snippet.
Each query is represented as an automaton executed using a finite state machine (FSM) in
an oblivious manner. Communication complexity is
$O(n\cdot(|\Sigma|+|Q|))$, where $n$ is snippet length, $|\Sigma|$ -- alphabet size (i.e., 4 for DNA),
and $|Q|$ -- number of states.  Computational complexity is $O(n\cdot|\Sigma|\cdot|Q|)$ and $O(n\cdot|Q|)$
cryptographic operations for Alice and Bob, respectively.
However,  the number of FSM states is always revealed to all parties.
To obtain error-resilient and approximate DNA matching,~\cite{ccs07} also shows how to 
construct an automaton that, given Alice's string $x$, accepts
all strings with Levenshtein distance~\cite{levenshtein} at most $d$ from $x$.

Blanton and Aliasgari~\cite{blanton_dbsec10} improve on~\cite{ccs07}, reducing Alice's work by a
factor of $|\Sigma|$ and Bob's --- by a factor of $\log(|Q|)$, incurring, however, a potentially
increased communication complexity (if the security parameter is smaller than $\log(|Q|)$).
This work also introduces a protocol for secure outsourcing  of
computation to an external service provider and a modified {\em multi-party} protocol.

A set of cryptographic protocols for secure pattern matching are presented in~\cite{genn-pkc10}
and~\cite{hazayAC10}.  Given a binary string $T$ of length $n$, held by Alice, and a binary pattern
$p$ of length $m$, held by Bob, pattern matching lets Bob learn all locations in $T$ where
$p$ appears. Secure computation guarantees that nothing except $m$ is learned by Alice,
and nothing about $T$ is revealed to Bob (besides $n$ and locations where $p$ appears).
\cite{genn-pkc10} proposes one such protocol, secure in the semi-honest setting, based on 
homomorphic encryption, with $O(m+n)$ communication and computation complexities. It 
includes another protocol, secure in the malicious setting, based on secure oblivious
automata evaluation, with quadratic complexity and $m$ rounds.
Subsequently, \cite{hazayAC10} presented an improved protocol, with malicious security,
using homomorphic encryption and incurring $O(m+n)$ complexity.

Another related result is the recent work in~\cite{Katz-ccs10}. It
realizes secure computation of the CODIS test~\cite{CODIS}
(run by the FBI for DNA identity testing), that could not be implemented using pattern matching or FSM.
It achieves efficient secure computation of function $M(T,p,e,l) \equal 1$ iff $|l_{max}(T,p)-l|\leq\varepsilon$,
where $T$ is a DNA fragment, $p$ a pattern, $(\varepsilon,l)$ some additional information, and
$l_{max}(T,p) \geq 0$ is the largest integer $l'$ for which $p^{l'}$ appears as a substring in $T$.
A general technique for secure text processing is introduced, combining garbled circuits and secure
pattern matching. (The latter is reduced to private keyword search and solved using
Oblivious Pseudorandom Functions (OPRF-s)~\cite{TCC05,HL08}.)
The resulting protocol can compute several functions (including CODIS) on sample $T$ and
pattern $p$, using the number of circuits linear in the number of
occurrences of $p$. Complexity incurred by the underlying keyword search protocol
is linear in $|T|$. However, common knowledge of some threshold on the number of occurrences
needs to be assumed.

\subsubsection*{Similarity of DNA Sequences}
Another set of cryptographic results focus on privately computing the {\em edit distance} of two
strings $\alpha,\beta$ of size $m$ and $n$, respectively.\footnote{Edit distance is the minimum number of operations
(delete, insert, or replace)  needed to transform $\alpha$ into $\beta$.} Privacy-preserving computation of
Smith-Waterman scores \cite{smith-waterman} has also been investigated and used for sequence alignment.

Jha, et al.~\cite{jha2008} proposed techniques for secure edit distance using
garbled circuits~\cite{yao}, and showed that the overhead is acceptable only for 
small strings (e.g., a 200-character strings require 2GB circuits). For longer strings, 
two optimized techniques were proposed; they exploit
the structure of the dynamic programming problem (intrinsic to the specific circuit)
and split the computation into smaller component circuits.
However, a quadratic number of oblivious transfers is needed to evaluate
garbled circuits, thus limiting scalability of this approach. 
For example, 500-character string instances take almost one hour to complete~\cite{jha2008}.
Optimized protocols also extend to privacy-preserving Smith-Waterman scores~\cite{smith-waterman},
a more sophisticated string comparison algorithm, where costs of delete/insert/replace operations,
instead of being equal, are determined by special functions.  Again, scalability is limited: experiments
in~\cite{jha2008} show that evaluation of Smith-Waterman for a 60-character string takes about
1,000 seconds.

Somewhat less related techniques include \cite{kanta2008} that proposed
a cryptographic framework for executing queries on genomic da\-ta\-bases where privacy
is attained by relying on two anonymizing and non-colluding parties.
Danezis, et al.~\cite{danezis2007} used negative databases to test a single profile against a database
of suspects, such that database contents cannot be efficiently enumerated.

\subsubsection*{Specialized Protocols}
Wang, et al.~\cite{wang_ccs09} proposed techniques for computation on genomic data stored at a data provider,
including:  edit distance, Smith-Waterman and search for homologous genes.
Program specialization is used to partition genomic data into ``public'' (most of the genome) and
``sensitive'' (a very small subset of the genome). Sensitive regions are replaced with symbols by
data providers (DPs) before data consumers (DCs) have access to genomic information.
DCs perform concrete execution on public data and symbolic execution on sensitive data, and may
perform queries to DPs on sensitive nucleotides. However, only queries that do not let DCs reconstruct
sensitive regions are  allowed by DPs and generic two-party computation techniques are used during query execution.
Portions of sensitive data are public information. We note that, due to the current limited knowledge of
human genome, parts that are considered non-sensitive today may actually become sensitive later. 

Also, Bruekers, et al.~\cite{eprint08} presented privacy-preserving techniques for a few
DNA operations,  such as: identity test, common ancestor
and paternity test, based on STR (Short Tandem Repeat; see Sec.~\ref{sec:bio-back}).
Homomorphic encryption is used on alleles (fragments of DNA) to compute  comparisons.
Testing protocols tolerate a small number of errors, however,
their complexity increases with the number of tolerated errors~\cite{blanton_dbsec10}.
Also,~\cite{eprint08} leaves as an open problem the scenario where an attacker (honestly) runs the protocol but
executes it on arbitrarily chosen inputs. In this setting, attackers, given STR's limited entropy,
can ``lie'' about their STR profiles and run multiple dependent protocols thus 
reconstructing the other party's profile.

\subsubsection*{Using Current Techniques?}
We aim to obtain secure and private computation on \FSGS{},
in scenarios where individuals possess their own genomic data.
As discussed in~Sec.~\ref{sec:intro}, we focus on paternity testing,
personalized medicine and genetic compatibility testing. Prior work
has yielded a number of elegant (if not always efficient) cryptographic protocols
for secure computation on DNA sequences. However, 
we identify some notable open problems:\vspace{0.1cm}
\begin{compactenum}
\item \textbf{\em Efficiency:} Most current protocols are designed for DNA snippets
(e.g., hundreds of thousands nucleotides) and it is unclear how to scale them
to full genomes (i.e., three billion nucleotides).
\item \textbf{\em Error Resilience:} Most prior work attempts to \linebreak achieve resilience to sequencing
errors {\em in computation} (e.g., using  approximate matching or distance with errors).
Not surprisingly, this results in: (i) significant computation and communication
overhead, and (ii) ruling out more efficient and simpler cryptographic tools, i.e., those geared for
exact matching. (Whereas, our goal is error-resilience by design.)
Also, as the cost of \FGS{} drops, so do error rates. By increasing the number of sequencing runs,
the probability of sequencing errors can be rapidly reduced.
\item \textbf{\em Inter-String Distance:} Analyzing the distance between sequenced
strings works for the creation of phylogenetic trees, parental analysis, and homology
studies.  However, it does not suit applications, such as genetic diseases testing,
that require much more complex comparisons.
\item \textbf{\em Paternity Testing:} To the best of our knowledge, the only available
technique for privacy-preserving genetic paternity testing is~\cite{eprint08}. However,
it does not prevent a participant from 
manipulating its input to reconstruct the counterpart's
profile. Also, as shown in Sec.~\ref{sec:paternity}, overhead can be significantly reduced using techniques
that obtain error resilience by design.
\item \textbf{\em Genetic Testing via Pattern Matching:} The use of  pattern matching over full genomes
to test for genetic
compatibility and/or personalized medicine is not straightforward. Suppose that a party wants to privately search
for certain gene mutation, e.g., Beta-Thalassemia. The pattern representing this
mutation might be very short --- a few nucleotides --- but needs to be searched in the full genome,
as restricting the search to the specific gene would trivially expose the nature of the test.
Therefore, \naive\ application of pattern matching would return all locations (presumably millions) where the pattern appears.
This would be detrimental to both privacy and efficiency of the resulting solution. We could modify the pattern to include nucleotides
expected to appear immediately before/after the mutation, such that, with high probability, this pattern
would appear at most once. However, this needs to be done carefully, since: (i) nucleotides added to the pattern
must appear in {\em all} human genomes, and (ii) the choice of pattern length should not expose the
mutation being searched. Plus, extending the pattern would also increase computation and communication overhead.
\end{compactenum}

\section{Preliminaries} \label{prelim}
This section provides some relevant biology and cryptography background information.

\subsection{Biology Background}\label{sec:bio-back}

\descr{Genomes} represent the entirety of an organism's hereditary information.
They are encoded either in DNA or, for many types of viruses, in RNA. The genome includes 
both the genes and the non-coding sequences of the DNA/RNA. For humans and many other organisms, 
the genome is encoded in double stranded deoxyribonucleic acid (DNA) molecules, 
consisting of two long and complementary polymer chains of four simple units called nucleotides, 
represented by the letters A, C, G, and T. The human genome consists of approximately 3 billion letters.

\descr{Restriction Fragment Length Polymorphisms (RFLPs)} refers to a difference between samples of 
homologous DNA molecules that come from differing locations of restriction enzyme sites, and to a 
related laboratory technique by which these segments can be illustrated. In RFLP analysis, a DNA 
sample is broken into pieces (digested) by restriction enzymes and the resulting restriction fragments 
are separated according to their lengths by gel electrophoresis. Thus, RFLP provides information about 
the length (but not the composition) of DNA subsequences occurring between known subsequences  
recognized by particular enzymes. Although it is being progressively superseded by inexpensive DNA 
sequencing technologies, RFLP analysis was the first DNA profiling technique inexpensive enough for 
widespread application. It is still widely used at present.
RFLP probes are frequently used in genome mapping and in variation analysis, such genotyping, 
forensics, paternity  tests and hereditary disease diagnostics. (For more details, see~\cite{RFLP}.)

\descr{Single Nucleotide Polymorphisms (SNPs)} are the most common form of DNA variation occurring
when a single nucleotide (A, C, G, or T) differs between members of the same species or paired chromosomes 
of an individual~\cite{stenson}. The average SNP frequency in the human genome is approximately 1 per
1,000 nucleotide pairs.\footnote{NCBI maintains an interactive collection of SNPs, dbSNP, containing all known 
genetic variations of the human genome~\cite{dbSNP}.}
SNP variations are often associated with how individuals develop diseases and respond to pathogens,
chemicals, drugs, vaccines, and other agents. Thus SNPs are
key enablers in realizing {\em personalized medicine}~\cite{Carlson}. Moreover, they are used in genetic
disease and disorder testing, as well as to compare genome regions between cohorts in genome-wide 
association studies.

\descr{Short Tandem Repeats (STRs)} occur when a pattern of two or more nucleotides are repeated and
repeated sequences are directly adjacent to each other. The pattern can range in length from 2 to 50 
nucleotides or so. Unrelated people likely have different numbers of repeat units in highly polymorphic 
regions, hence, STRs are often used to differentiate  between individuals. STR {\em loci} (i.e., locations 
on a chromosome) are targeted with sequence-specific primers. 
Resulting DNA fragments are then separated and detected using electrophoresis.
By identifying repeats of a specific sequence at specific locations in the genome, it is possible to create 
a genetic profile of an individual. There are currently over 10,000 published STR sequences 
in the human genome.

\subsection{Cryptography Background}
We now overview a set of cryptographic concepts and tools used in the rest of the paper.
For ease of exposition, we omit basic notions and refer to~\cite{Goldreich,katz-lindell,Applied}
for details on various cryptographic primitives, such as hash functions, number-theoretic assumptions,
as well as encryption and signature schemes.

\descr{Private Set Intersection (PSI)} \cite{FNP}:
a protocol between \Server{} with input $\mathcal{S} \equal \{s_1,\ldots,s_w\}$,
and \Client{} with input $\mathcal{C} \equal \{c_1,\ldots,c_v\}$. At the end,
\Client{} learns $\mathcal{S}\cap\mathcal{C}$.
PSI securely implements:
$\mathcal{F}_{\sf PSI}: (\mathcal{S}, \mathcal{C}) \mapsto (\perp,\mathcal{S} \cap \mathcal{C}).$

\descr{Private Set Intersection Cardinality (PSI-CA)} \cite{FNP}:
a protocol between \Server{} with input $\mathcal{S} \equal \{s_1,\ldots,s_w\}$,
and \Client{} with input $\mathcal{C} \equal \{c_1,\ldots,c_v\}$. At the end,
\Client{} learns $|\mathcal{S}\cap\mathcal{C}|$.
PSI-CA securely implements:
$\mathcal{F}_{\sf PSI\mbox{-}CA}: (\mathcal{S}, \mathcal{C}) \mapsto (\perp, \linebreak |\mathcal{S} \cap \mathcal{C}|).$

\descr{Authorized Private Set Intersection (APSI)} \cite{FC10}:
a protocol between \Server{} with input $\mathcal{S} \equal \{s_1,\ldots,s_w\}$,
and \Client{} with input $\mathcal{C} \equal \{c_1,\ldots,c_v\}$ and $\mathcal{C}_\sigma \equal \{\sigma_1, \ldots, \sigma_v\}$.
At the end, \Client{} learns:
$$\mbox{\sf \small ASI} \hspace{-0.05cm}\stackrel{\text{\tiny def}}{=}\hspace{-0.05cm}
\mathcal{S}\cap \{c_i \; | \; c_i\in \mathcal{C} \; \wedge \; \sigma_i \mbox{ \emph{valid auth. }}
\mbox{\em  on } c_i\}.$$
APSI securely implements:
$\mathcal{F}_{\sf APSI}: (\mathcal{S}, (\mathcal{C},\mathcal{C}_\sigma)) \mapsto (\perp,\mbox{\sf \small ASI}).$

\begin{figure*}[t!]
\centering
\fbox{\small
\begin{minipage}{1.75\columnwidth}
\begin{tabular}{lcl}
\textbf{\underline{\Client}}, on input \spa $\Cset=\{c_1,\ldots,c_v\}$ &\hspace{-0.25cm}[Common Input: $(p,q,g,\mbox{H},\mbox{H}')$] &  \textbf{\underline{\Server}}, on input \spa  $\Sset=\{ s_1,\ldots,  s_w\}$ \vspace{0.2cm}\\
&& \underline{\bf Offline} \vspace{0cm}\\
& & $\{\hat s_1,\ldots, \hat s_w\} \leftarrow \Pi (\Sset)$, with $\Pi$   \\
& &  random permutation   \\
&&$R_s \leftarrow \Z_{q},\spa R_s' \leftarrow \Z_{q}$, $Y=g^{R_s}$\\
&&$\forall j\spa 1\leq j\leq w,\spa {ks_{j}} = \mbox{H}(\hat s_j)^{R_s'}$\\
&\\
\underline{\bf Offline} && \\
$R_c \leftarrow \Z_q, \spa R_c' \leftarrow \Z_q $, $X=g^{R_c}$ & \hspace{-0.6cm} $X, \{a_1,\ldots,a_v\}$ & \underline{\bf Online}\\
$\forall i\spa 1\leq i\leq v, \spa a_i={\mbox{H}(c_i)}^{R_c'}$ &\hspace{-0.6cm}  $\xymatrix@1@=85pt{\ar[r]^*+<4pt>{}&}$ &$\forall i\spa 1\leq i\leq v,\spa a_i' =(a_i)^{R_s'} $\vspace{0.1cm}\\
&& $(a'_{\ell_1}, \ldots, a'_{\ell_v}) = \Pi(a'_{1},\ldots,a'_{v})$\vspace{0.1cm}\\
&& $\forall j\spa 1\leq j\leq w,\spa ts_{j} =\mbox{H}'(X^{R_s}\hspace{-0.08cm}\cdot{ks_{j}})$\vspace{-0.6cm}\\
\underline{\bf Online} &\hspace{-0.6cm}  $\xymatrix@1@=85pt{& \ar[l]^*
+<4pt>{\{ts_{1},\ldots,ts_{w} \}}_*+<4pt>{Y,\{a'_{\ell_1},\ldots,a'_{\ell_v}\}}}$\vspace{-0.4cm}\\
$\forall i\spa 1\leq i\leq v,\spa tc_{\ell_i} =\mbox{H}'((Y^{R_c})(a'_{\ell_i})^{1/R_c'})$\vspace{0.1cm}\\
{\bf Out}: $\left| \{ts_{1},\ldots,ts_{w}\}  \cap \{tc_{\ell_1},\ldots,tc_{\ell_v}\} \right|$&&\\
\end{tabular}
\end{minipage}
}
\caption{PSI-CA protocol from~\cite{ePrint}. It executes on common input of two primes $p$ and 
$q$  (such that $q|p-1$), a generator $g$ of a subgroup of size $q$ and two hash functions, H and H$'$, modeled as random oracles.  All computation is  mod $p$.}
\vspace{-0.2cm}
\label{fig:PSI-CA}
\end{figure*}
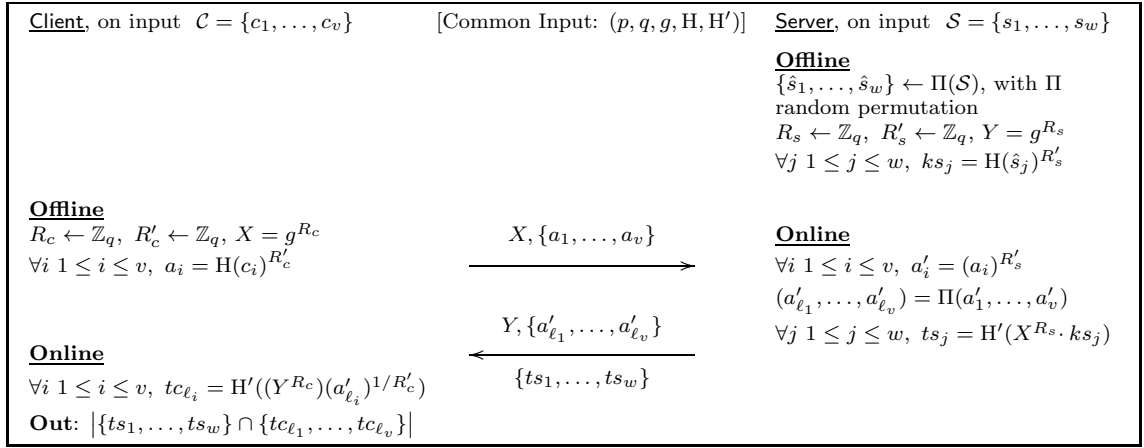

\descr{Additively Homomorphic Encryption.} Let $(K;Enc;Dec)$ be a homomorphic encryption scheme,
where $K$ is the key generator algorithm selecting public/secret key-pair $(pk,sk)$.
Assume that the message space for a public key $pk$ is $\Z_p$ for some integer $p$,
then $Enc(m)$ denotes encryption under key $pk$, and $Dec(c)$ denotes decryption under
key $sk$.
The following additive homomorphic properties hold:
(1) the product of two ciphertexts is a ciphertext for the sum of the plaintexts, i.e.,
for any $a,b\in\Z_p$, we have $Dec(Enc(a)\cdot Enc(b))=a+b$,
and (2) raising a ciphertext for a message $a$ to power $r$ gives a ciphertext $r\cdot a$, i.e.,
for any $r\in\Z_p$, we have $Dec(Enc(a)^r)=r\cdot a$.

\descr{Adversarial Model.} 
We use standard security models for secure two-party computation.
One distinguishing factor is  the adversarial model that is either semi-honest or malicious.
(In the rest of this paper, the term {\em adversary} refers to insiders, i.e., protocol participants. Outside adversaries
are not considered, since their actions can be mitigated via standard network security techniques.)

Following definitions in~\cite{Goldreich}, protocols secure in the presence of
{\em semi-honest adversaries} assume that parties faithfully follow
all protocol specifications and do not misrepresent any information related to their inputs,
e.g., size and content. However, during or after protocol execution, any party might
(passively) attempt to infer additional information about the other party's input.
This model is formalized by considering an ideal implementation where a trusted third party (TTP)
receives the inputs of both parties and outputs the result of the defined function. Security in the presence
of semi-honest adversaries requires that, in the real implementation of the protocol (without a TTP),
each party does not learn more information than in the ideal implementation.

Security in the presence of {\em malicious parties} allows  arbitrary  deviations
from the protocol. However, it does not prevent  parties from refusing to participate
in the protocol, modifying their inputs, or prematurely aborting the protocol.
Security in the malicious model is achieved if the adversary (interacting in the real protocol,
without the TTP) can learn no more information than it could in the ideal scenario. In other words,
a secure protocol emulates (in its real execution) the ideal execution that includes a TTP.
This notion is formulated by requiring the existence of adversaries in the ideal execution model that
can simulate adversarial behavior in the real execution model.

Although security arguments in this paper are made with respect to semi-honest participants,
extensions to malicious participant security (with the same computation and communication complexities) 
have already been developed for our cryptographic building blocks: PSI, PSI-CA and APSI. We
consider these extensions to be out of the scope of this paper.

\section{ 
Genome Testing}\label{sec:our}
We now explore efficient techniques for privacy-preserving testing on \FSGS{}. Unlike most
prior work (reviewed in Sec.~\ref{sec:related}), we do not seek generic solutions for genomic 
computation. Instead, we focus on a few specific real-world applications and, for each, 
capitalize on domain knowledge to propose an efficient privacy-preserving approach.

\descr{Notation.} We assume that each participant has a digital copy of her \FSG{}
denoted by $\mathcal{G} \equal \{(b_1||1),\ldots,(b_n||n)\}$, where $b_i \in \{\mbox{A, G, C, T, --}\}$,
 $n$ is the human genome length (i.e., $3\cdot10^9$), and ``$||$'' denotes concatenation.
The ``--'' symbol is needed to handle DNA mutations corresponding to {\em deletion}, i.e., where 
a portion of a chromosome is missing~\cite{lewis2003human}. It is also used
when the sequencing process fails to determine a nucleotide.
This data may be pre-processed in order to speed up execution of specific applications.
For example, parties may pre-compute a cryptographic hash, H$(\cdot)$, on each nucleotide, alongside 
its position in the genome, i.e., for each  $(b_i||i) \in \mathcal{G}$, they compute 
$hb_i \equal \mbox{H}(b_i||i)$.\footnote{In case of {\em insertion} mutation in the genome, e.g., an `A' is
added between positions 35 and 36, genome pre-processing computes H$($A$||35||1$).  
Similarly, if insertion involves multiple nucleotides. Since insertions are rare in human genomes,
we do not consider them in this paper.}

We use the notation $|str|$ to denote the length of string $str$,
and $|A|$ to denote the cardinality of set $A$. Finally,
we use $r \leftarrow R$ to indicate that $r$ is chosen
uniformly at random from set $R$.

\descr{Experimental Setup.} The rest of this section includes some experimental results.
Unless explicitly stated otherwise, all experiments were performed on a Linux Desktop,
with an Intel Core i5-560M (running at 2.66 GHz). All tests were run on a single processor core
and all code is written in C, using OpenSSL and GMP libraries.
Cryptographic protocols use the SHA-1 hash function and 1024-bit moduli.
Source code of our experiments is available at  \url{http://sprout.ics.uci.edu/projects/privacy-dna}.

\subsection{Genetic Paternity Test}\label{sec:paternity}
A Genetic Paternity Test (GPT) allows two individuals with their respective  genomes
to determine whether there exists a
biological parent-child relationship between them. A \textbf{\em Privacy-Preserving Genetic Paternity Test} (PPGPT)
achieves the same result without revealing any information about the two genomes. In the following,
we refer to the two participants as \Client\ and \Server. Only \Client\ receives the outcome of the test.

\subsubsection{Strawman Approach}\label{sec:strawman}
Genomics studies have shown that about 99.5\% of any two human genomes are identical.
Humans carry two copies of each chromosome, inherited one from the mother and one from the father.
Thus, genomes carried by two individuals tied by a parent-child relationship show an even higher degree of similarity.
As a result, one immediate computational technique for GPT is to compare the candidate's genome with that of the child;
the test returns a positive result if  the percentage of matching nucleotides is above a given threshold $\tau$,
i.e., significantly higher than 99.5\%.

\descr{First-Attempt Protocol.} At first glance, protecting privacy
is relatively easy: recent proposals for Private Set Intersection Cardinality
(PSI-CA) protocols \cite{FNP,vaidya2005,kissner2005,ePrint} offer
efficient and private two-party computation of the number of set elements
shared by two parties. Thus, to perform PPGPT, two participants just need to
run PSI-CA on input of their respective genomes.

We select the PSI-CA construction from~\cite{ePrint} (shown in Fig.~\ref{fig:PSI-CA})
since 
it offers the best communication and computation complexities.
Also, we use PSI-CA rather than PSI since {\em semi-honest} participants only need to learn 
{\em how similar} their genomes are. Whereas, PSI would also reveal  {\em where} the two genomes 
differ and/or where they have common features.

We emphasize that this approach provides very accurate results, and is
not significantly affected by potential sequencing errors. In fact, given 
expected error ratio $\varepsilon$, one can simply
modify threshold $\tau$ to accommodate errors. This is because $\varepsilon$ 
is expected to be significantly smaller than the difference between $\tau$ and the percentage 
of nucleotides that any two individuals share.

Unfortunately, since the number of nucleotides in the human genome is extremely large (about $3\cdot10^9$),
this technique, though optimal in terms of accuracy, is impractical using current commodity hardware, as it requires
both parties to perform online computation over the entire genome.  Specifically, PSI-CA
entails a number of (short) modular exponentiations linear in the input size.
Table~\ref{tab:paternity1} estimates execution times and bandwidth incurred by this na\"ive 
approach. Since \Client's online computation depends on that of the \Server, a single test would consume
approximately 10 days.

\begin{table}[htbp]
\vspace{-0.1cm}
\centering
\begin{tabular}{|l|c|c|c|c|}
\cline{2-4}
\multicolumn{1}{c|}{}
& \multicolumn{1}{c|}{Offline} & \multicolumn{2}{c|}{Online} \\ \cline{3-4}
\multicolumn{1}{c|}{}
& Time  & Time & Size \\
\hline
\Client{} & $4.5$ days & $4.5$ days & $358$ GB\\
\hline
\Server{} & $4.5$ days  & $4.5$ days & $414$ GB\\
\hline
\end{tabular}
\vspace{-0.2cm}
\caption{Computation and communication costs of the first straw-man PPGPT protocol.} 
\label{tab:paternity1}
\vspace{-0.2cm}
\end{table}

\descr{Improved Protocol.\hspace{0.1cm}}
Since about 99.5\% of the human genome is the same, two parties would only need to
compare the remaining 0.5\%. Unfortunately, there is yet not enough statistical
knowledge to pinpoint {\em where} exactly this 0.5\% occurs. Nonetheless, experts claim that,
in practice, comparing a properly chosen 1\% of the genome yields an accuracy comparable
to analyzing the entire genome \cite{gibbssingleton06}. Running times and bandwidth overhead
required by this improved method are presented in Table~\ref{tab:paternity2}.

\begin{table}[htbp]
   \centering
\begin{tabular}{|l|c|c|c|c|}

\cline{2-4}
\multicolumn{1}{c|}{}
& \multicolumn{1}{c|}{Offline} & \multicolumn{2}{c|}{Online} \\ \cline{3-4}
\multicolumn{1}{c|}{}
& Time  & Time & Size \\
\hline
\Client{} & $67$ mins &  $67$ mins & $3.57$ GB\\
\hline
\Server{} & $67$ mins &  $67$ mins & $4.14$ GB\\
\hline
\end{tabular}
\caption{Computation and communication costs of improved PPGPT protocol.
Computation is performed over 1\% of the human genome.}
\label{tab:paternity2}
\vspace{-0.0cm}
\end{table}

\subsubsection{Efficient RFLP-based PPGPT with PSI-CA}\label{sec:rflp}
We now present a very efficient technique for Privacy-Preserving Genetic Paternity Testing (PPGPT).
To construct it, we take advantage of domain knowledge in genomics and build upon effective
{\em in vitro} techniques (RFLP or SNP) rather than generic computational techniques.  First, we design a
protocol that implements  RFLP-based GPT. Next, we propose a cryptographic technique for secure
computation of this protocol that realizes PPGPT. Finally, we show that the technique used for computing 
RFLP-based GPT can be easily adapted to perform SNP-based GPT.

As discussed in Sec.~\ref{sec:bio-back}, RFLPs use specific restriction enzymes (e.g., HaeIII, PstI, and HinfI),
to digest a genome into hundreds of  smaller fragments. Following the deterministic and well-known process,
enzymes cut the DNA at each occurrence of a given pattern (e.g., ``{\tt CTGCAG}'' with PstI).
Next, a subset of these fragments is selected using a small number of probes for well-known
markers, which are located in known areas of the genome. In an RFLP-based paternity test, this process is applied to
the DNA of the two tested individuals. If resulting fragments have comparable lengths, then the test returns a positive with certain confidence, based on the exact number of fragments
of the same length.

There are a few slightly different ways to select the type and the number of markers, thus identifying
exactly which fragments to compare. For the sake of reliability,  one needs to use markers that are rare
enough (i.e., occur in unrelated individuals with very low probability) while
common enough to occur in at least one of the tested subjects.
Currently, public databases and scientific literature offer thousands available probes for RFLP in
human genomes \cite{canada,applmapping,tokino1991}. However, to reduce the cost of {in vitro} tests,
only a small subset of them is actually used \cite{dracopoli1994}. Different laboratories consider various
accuracy/cost trade-offs. Some compare as few as 9-15 DNA markers, returning a positive result whenever
fewer than two fragments do not match~\cite{dailybusiness03}, with an estimated 99.9\%
accuracy. Meanwhile, others use up to 25 markers
and return a positive whenever fewer than two fragments do not match,
thus providing significantly higher accuracy, i.e., about 99.999\%~\cite{endeanrflp,lander1989dna}.

In the United States, these testing methodologies follow precise regulations issued by the American Association of
Blood Banks (AABB) and are considered legally admissible as evidence in the court of law.
Since our PPGPT technique closely mimics the {\em in vitro} procedure, it achieves the same level of accuracy.
Nevertheless, as the cost of RFLP emulation on digitalized genomes is not significantly affected by
the number of selected markers, we can anticipate increasing the number of markers to improve accuracy.
We could perform tests with 50 markers and show that this only adds a small cost. However,
selection of additional markers is out of the scope of this paper, as their introduction does
not change the algorithm's functionality presented below.

\descr{RFLP-based Protocol.} This  protocol
involves two individuals, on private input of  their respective \FSGS. We distinguish
between \Client{} and \Server{}, to denote the fact that only the former learns the test outcome.
The protocol is run on common input of: a threshold $\tau$, a set of enzymes $E \equal \{e_1, \ldots, e_j \}$,
and a set of markers $M \equal \{ mk_1, \ldots , mk_l \}$.
Each participant also inputs its digitized genome.
\begin{compactenum}
\item First, participants emulate the digestion process of each enzyme $e_i \in E$ on their genome.
Consider, for instance, the PstI enzyme: whenever the string {\tt CTGCAG} occurs, the enzyme  cuts
the genome in two fragments, so that the first ends with {\tt CTGCA} and the second starts with {\tt G}.
As a result, genomes are digested into a large number of fragments of variable length.
\item Next, participants probe the fragments using markers in $M$. During this process, each participant
selects up to $l$ fragments $\{frag_1,\ldots,frag_l\}$ (e.g., $l\equal25$), corresponding to $M$. All
remaining fragments are discarded. Public markers are chosen such that each
appears in at most one sequence.
\item \Client{} builds the set $F_C\equal\{(|frag^{(c)}_{i}|, mk_i)\}_{i=1}^l$. For each marker $i$ not
corresponding to any fragment, $frag^{(c)}_{i}$ is replaced with the empty string. Similarly, \Server{}
builds $F_S \equal \{(|frag^{(s)}_{i}|, mk_i)\}_{i=1}^l$ 
\item \Client{} and \Server{} run the PSI-CA protocol described in Fig.~\ref{fig:PSI-CA}, on respective inputs:
$F_C$  and $F_S$. \Client{} learns $pt \equal |F_C\cap F_S|$, i.e., how many of its and \Server's fragments are
of the same size.
\item \Client{} learns the test result by comparing $pt$ to threshold
$\tau$.
\end{compactenum}

\descr{Why Compare Lengths?}
It might seem that comparing string {\em lengths} is unreliable since two same-length strings might encode
completely different content, while our protocol would consider these strings as matching.
In practice, however, this well-established technique yields false positives with \textbf{\em extremely low probability}. 
Sequences are selected using markers, i.e., according to (part of) their content. Selection of markers, in turn, guarantees
that they appear only in one specific position in the entire genome.  Edges of each fragment are content-dependent
as well, since enzymes digest them according to a specific pattern of nucleotides. Therefore, two unrelated
sequences of the same length would not be compared and two same-length sequences containing
the same marker should be indeed considered matching.

Furthermore, this approach boosts the resilience of PPGPT a\-gainst sequencing errors.
Only errors occurring in the pattern digested by enzymes (or in the markers) influence
the result of the RFLP-based PPGPT. However, since patterns and markers are relatively short
compared to the size of the genome, this happens with very low probability, since sampling errors
are uniformly distributed. However, if we let participants compare hashes of fragments, rather
than their length, even a moderate error rate would severely increase the probability of
false negatives, since even a single sequencing error would affect the final outcome of the test.
Moreover, the main purpose of the PPGPT presented in this paper is not to improve accuracy
of the {\em in vitro} test currently used, but to efficiently and securely replicate it {\em in silico}.

\descr{PSI-CA or PSI?}
The use of PSI-CA, rather than PSI, is needed to minimize information learned by \Client{} from  protocol
execution.
With PSI, if the number of matches is sufficiently high
(even if the test is negative), \Client{} would learn the lengths of several \Server's
fragments: it could then use this information to perform a paternity test between the party
previously playing the role of \Server{} and any other individual (although with slightly lower reliability).

\descr{SNP-based Protocol.} SNP-based tests are replacing  RFLP-based tests due to their better performance 
\cite{SNPID2008}. While this technique is not yet considered legally admissible in court, it is expected to eventually 
supersede its RFLP-based counterpart. Our RFLP-based protocol can be extended to perform paternity testing using 
SNPs: instead of selecting fragments using enzymes and markers, the SNP-based test  selects fragments using a 
set of known SNPs. Since the rest of the protocol is unchanged and the size of the set of SNPs is usually 
52 elements~\cite{SNPID2008}, the new protocol performs almost identically to the RFLP-based PPGPT 
protocol with 50 fragments.

\descr{Performance Evaluation.} We now measure performance of the RFLP-based protocol on the 
Intel Core i5-560M testbed.  The (offline) time needed to emulate the enzyme digestion process on the full 
genome is 74 seconds. This computation is performed only once, thus, it does not affect the time required 
to perform the interactive protocol. Finally, in order to assess the practicality of the protocol on embedded 
devices,  we also measured its performance on a modern smartphone --- a
Nokia N900 equipped with ARM Cortex A8 CPU running at 600 MHz.
Table~\ref{tab:paternity3} summarizes the online cost of the RFLP-based protocol, measuring computation and
communication overhead, using different numbers of markers, on both i5-560M and A8 processors.

\begin{table}[htbp]
\centering {\small
\begin{tabular}{|l|c|c|c|c|c|}
\hline
{ Entity} & \multicolumn{2}{c|}{Offline (Time)} & \multicolumn{3}{c|}{Online (Time/size)} \\
 (markers)  & i5-560M & A8  & i5-560M & A8 & Size \\
\hline
{\sf  Client} (25) & 3.4 ms & 323 ms  & 3.4 ms & 323 ms  & 3 KB\\
\hline
{\sf  Server} (25) & 3.4 ms & 323 ms  & 3.4 ms & 323 ms & 3.5 KB\\
\hline
{\sf  Client} (50) & 6.7 ms & 645 ms  & 6.7 ms & 645 ms  & 6 KB\\
\hline
{\sf  Server} (50) & 6.7 ms & 645 ms  & 6.7 ms & 645 ms  & 7 KB\\
\hline
\end{tabular}
\vspace{-0.15cm}
\caption{Computation and communication costs of RFLP-based PPGPT technique, testing 
25 and 50 fragments.}
\vspace{-0.2cm}
\label{tab:paternity3}}
\end{table}

For the sake of completeness, we compared our results to prior work on privacy-preserving paternity testing, presented in 
Figure 3 of~\cite{eprint08}. 
Following a conservative approach, we instantiate: (i) the cheapest protocol variant, which tolerates no error,
and (ii) the most efficient additively homomorphic cryptosystem among those suggested, 
i.e., modified ElGamal~\cite{elgamal}. Also, we only count the number of modular exponentiations.
Given that the paternity test is performed over $n$ alleles (with $n$ ranging from $13$ to $67$ for increasing accuracy) we 
estimate the following costs.
In step (2) of the protocol, the party obtaining the test result computes $8n$ modified ElGamal encryptions,
thus, incurring $24n$ (short) modular exponentiations. In the i5-560M testbed, this takes from $43$ms to $224$ms, 
depending on $n$. 
In step (3), the other party needs to obtain the encrypted sum using homomorphic properties: it does so by
performing $30n$ exponentiations. This takes between $54$ and $262$ms on the i5-560M testbed.
Even ignoring all other operations in~\cite{eprint08} and without pre-computation,
our most accurate test (using 50 markers) is about 5 times faster than the least accurate 
test in~\cite{eprint08} (using 13 alleles).

\subsubsection{PPGPT using Private Equality Testing}\label{sec:new}
We now discuss another approach to PPGPT that uses Private Equality Testing (PET) and 
homomorphic encryption. 

\descr{Comparing Two Genomes.} As mentioned in Section \ref{sec:strawman},
about $99.5\%$ of any two human genomes are identical. Therefore, the most natural
way of performing paternity test appears to be by determining how many nucleotides
are shared between them. 
We already outlined a mechanism for privacy-preserving computation 
of such a test by representing a genome as a (unordered) set where each element correspods to 
a position-numbered nucleotide and then using PSI-CA to obtain the number of matching nucleotides.
We now consider another technique. If we represent a genome as an ordered vector,
then we can count the number of matching nucleotides by testing pairwise equality
of vector elements. In the privacy-preserving ``world'', this problem 
can be solved using so-called Private Equality Testing (PET).

Given a probabilistic additively homomorphic encryption scheme,
$(K;Enc;Dec)$, such as modified ElGamal~\cite{elgamal},
Paillier~\cite{paillier1999public}, or DGK~\cite{dam08}, PET can be realized as follows. 
We assume \Client\ and \Server,
on input of two items $c$ and $s$, respectively, want to verify whether or 
not $c=s$:
\begin{enumerate}
\item \Client\ generates $(pk,sk)$ and sends \Server\ $Enc(c)$; 
\item \Server\ replies with $(Enc(c)\cdot Enc(-s))^r$, for some random $r$ in the message space; 
\item \Client\ learns that $c=s$ if \Server's answers decrypts to zero, and nothing otherwise;
\end{enumerate}
Note that ``$\cdot$'' denotes the operation on two ciphertexts
that results in the ciphertext of the sum of their plaintexts, e.g., $c$ and $-s$. Also, $Enc(x)^r$ is the 
operation on the ciphertext that yields modular exponentiation with exponent $r$
of the corresponding plaintext $x$.

By extending this PET to $n$ parallel executions (where $n$ is the
the genome size in nucleotides), \Client\ would learn how many
nucleotides they have in common. However, it would also learn {\em which} ones are shared.
To prevent the latter, we modify the protocol as follows. Assuming that $c_i$ and $s_i$
represent the $i$-th nucleotide in \Client's and \Server's genomes, respectively:
\begin{enumerate}
\item \Client\ generates $(pk,sk)$ and sends \Server:  
$$\{Enc(c_1),\ldots,\linebreak Enc(c_n)\}$$ 
\item \Server\ replies with: 
$$\Pi\left(\{(Enc(c_1)\cdot Enc(-s_1))^{r_1},\ldots,(Enc(c_n)\cdot Enc(-s_n))^{r_n}\}\right)$$ 
where $\Pi(\cdot)$ is a random permutation and $r_1,\ldots,r_n$ are random in the message space.
\item Using $\Pi(\cdot)$, \Client\ learns the number of ciphertexts that decrypt to yield a zero
and, yet, does not learn {\em which} nucleotides match.
\end{enumerate}

\descr{RFLP-based Paternity Test.} We can also use PET to obtain PPGPT based on RFLP.
Recall from Section~\ref{sec:rflp} that, after enzyme digestion and marker probing, \Client\ and \Server\
obtain, respectively, $\{|frag_i^{(c)|}\}_{i=1}^l$ and $\{|frag_i^{(s)}|\}_{i=1}^l$.
Similar to the technique discussed above, they can compute the number of fragments
of the same length by using PET. (Recall that a list of fragments is an ordered vector).
The resulting protocol is as follows: 
\begin{enumerate}
\item \Client\ generates $(pk,sk)$ and sends \Server: 
$$\{Enc(|frag_1^{(c)}|),\ldots, Enc(|frag_l^{(c)}|)\}$$
\item \Server\ replies with: 
\begin{eqnarray}
\nonumber
\Pi\left(\{(Enc(|frag_1^{(c)}|)\cdot Enc(-|frag_1^{(s)}|))^{r_1},\ldots,\right.& \\
\ldots,\left.(Enc(|frag_l^{(c)}|)\cdot Enc(-|frag_l^{(s)}|))^{r_l}\}\right)
\nonumber
\end{eqnarray}
where $\Pi(\cdot)$ is a random permutation and $r_1,\ldots,r_l$ are random in the message space.
\item By decryption, \Client\ learns the number of matching fragment lengths (and nothing else)
and determines the test outcome.
\end{enumerate}

\subsection{Personalized Medicine}\label{sec:PM}
Personalized Medicine (PM) 
is increasingly used to provide patients with drugs designed for their specific genetic features.
As discussed in Sec.~\ref{sec:intro}, in the context of PM, drugs are associated with a unique genetic fingerprint.
Their effectiveness is maximized in patients with a matching DNA~\cite{ho2010}.
To this end,  genomes need to be compared against the fingerprint and
a patient need to surrender her DNA to a physician or a pharmaceutical company.

One privacy-preserving approach is to let the patient independently run specialized
software over her genome and identify a match (or lack thereof) with a given drug's fingerprint. 
This way, the patient would learn whether the drug is appropriate.
However,  pharmaceuticals may consider DNA fingerprints of their drugs to be trade secrets
and thus might be unwilling to reveal them.
At the same time, for every new drug, pharmaceuticals are required to obtain approval from 
appropriate government entities, e.g., the Food and Drug Administration (FDA) in case of the
United States.

We now introduce a technique for \textbf{\em Privacy-Preserving Personalized Medicine Testing} (P$^3$MT),
involving the following steps:
\begin{compactitem}
\item Following positive clinical trials, a pharmaceutical company obtains
FDA approval on a specific DNA fingerprint $fp$ and receives a corresponding authorization, $auth$.
\item The pharmaceutical and the patient engage in a protocol, where the former inputs
$(fp,auth)$ and the latter inputs her genome.
\item At the end of the protocol, the pharmaceutical learns whether the patient's genome matches
fingerprint $fp$, provided that $auth$ is a valid authorization of $fp$.
\end{compactitem}
Privacy requirements are that: (1) the company learns nothing about
patient genome besides the part matching the (authorized) fingerprint, and (2) the patient learns
nothing about $fp$ or $auth$.

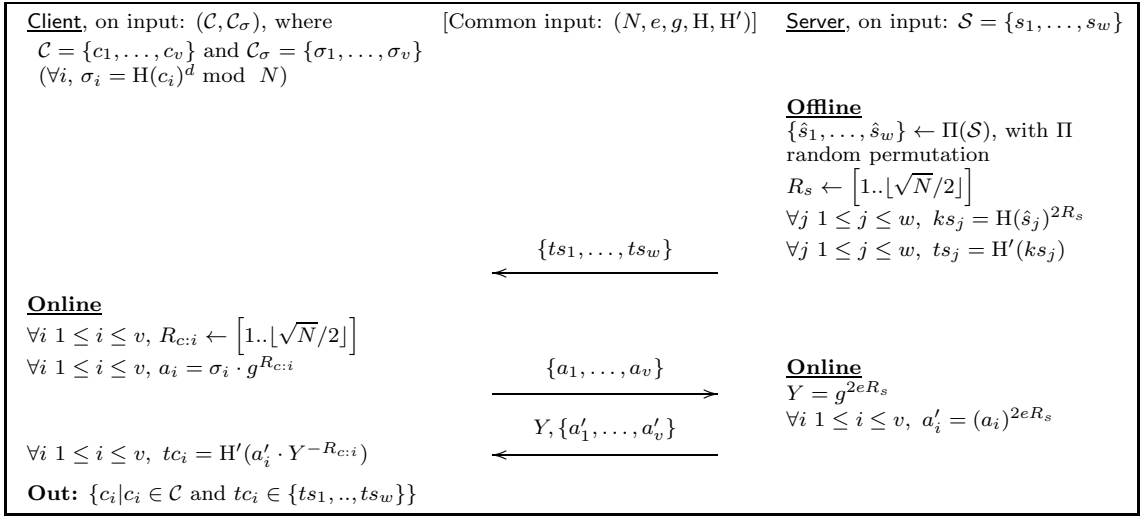
\begin{figure*}[t!]
\centering
\fbox{\small
\begin{minipage}{1.75\columnwidth}
\begin{tabular}{lcl}
\textbf{\underline{\Client}}, on input: $(\Cset,\Cset_\sigma)$, where \vspace{2mm} & \hspace{-0.23cm} [Common input:  $(N, e, g, \mbox{H}, \mbox{H}')$] & \textbf{\underline{\Server}}, on input: $\Sset=\{s_1,\ldots, s_w\}$ \vspace{-0.13cm} \\
\hspace{0.05cm} $\Cset=\{c_1,\ldots,c_v\}$ and $\Cset_\sigma=\{\sigma_1,\ldots,\sigma_v\}$ & & \\
\hspace{0.05cm} $(\forall i$, $\sigma_i = \mbox{H}(c_i)^d \bmod~N)$ \vspace{0.1cm}\\
 && \underline{\bf Offline}\\
& & $\{\hat s_1,\ldots, \hat s_w\} \leftarrow \Pi (\Sset)$, with $\Pi$  \\
&& random permutation\\
& & $R_s \leftarrow \left[1..\lfloor\sqrt{N}/2\rfloor\right]$\\
& &  $\forall j\spa 1\leq j \leq w, \spa ks_j = \mbox{H}(\hat s_j)^{2R_s}$\vspace{0.1cm}\\
&&  $\forall j\spa 1\leq j \leq w, \spa ts_{j} = \mbox{H}'(ks_j)$\vspace{-0.5cm}\\
 & $\xymatrix@1@=85pt{& \ar[l]^*
+<4pt>{}_*+<4pt>{\{ts_{1},\ldots,ts_{w} \}}}$\vspace{0.1cm}\\
\underline{\bf Online} & & \\
$\forall i\spa 1\leq i \leq v$, $R_{c:i} \leftarrow \left[1..\lfloor\sqrt{N}/2\rfloor\right] $ &  & \\
$\forall i\spa 1\leq i \leq v$, $a_i={\sigma_i}\cdot {g}^{R_{c:i}}$ & $ \{a_1, \ldots, a_v\} $ & \underline{\bf Online}\\
& $\xymatrix@1@=85pt{\ar[r]^*
+<4pt>{}_*+<4pt>{}&}$ & $Y=g^{2eR_s}$\\
&&$\forall i\spa 1\leq i \leq v, \spa  a'_i=(a_i)^{2eR_s}$\vspace{-0.2cm}\\
& $Y, \{a'_1, \ldots, a'_v\}$ &  \\
$\forall i\spa 1\leq i \leq v, \spa tc_i = \mbox{H}'(a'_i\cdot Y^{-R_{c:i}})$&  $\xymatrix@1@=85pt{& \ar[l]^*{}}$&\vspace{-0.3cm} \vspace{0.5cm}\\
 {\bf Out:}  $\{c_i | c_i \in \mathcal{C} $ and $tc_i \in \{ts_1,..,ts_w\} \} $&& \\

\end{tabular}
\end{minipage}
}
\vspace*{-0.1cm}
\caption{APSI Protocol from \cite{AC10} (simplified for semi-honest security). The protocol is run on common 
input of RSA modulus $N=pq$ (with $p$ and $q$ safe primes), public exponent $e$, a random element $g$ in 
$\mathbb{Z}^*_N$ and  two hash functions, H and H$'$, modeled as random oracles. All computation is mod $N$.}
\label{fig:apsi}
\end{figure*}

\subsubsection{P$^3$MT Instantiation}
We now present a specific P$^3$MT instantiation.
It involves: (1) an authorization authority (e.g., the FDA) denoted as \CA,
(2) a pharmaceutical --- \Client{}, and (3) a patient --- \Server.

Our cryptographic building block is Authorized Private Set Intersection (APSI)
\cite{FC09,FC10,AC10}, hence, our \Client/\Server/\linebreak\CA{} notation.
We select one specific APSI construction in~\cite{AC10}, illustrated in Fig.~\ref{fig:apsi},
since it currently offers lowest communication and computation complexity.
(Moreover, it can be instantiated in the malicious model with only a small constant additional overhead.) 
For efficiency reasons, $R_{c:i}$'s and $R_s$ are chosen uniformly at random from $W\equal[1..\lfloor \sqrt{N}/2 \rfloor]$, 
rather than from $\Z_{N/2}$, as in the original version of the protocol. In fact, 
as proved in~\cite{Goldreich00onthe}, the distribution of $g^x \bmod N$ with $x \leftarrow W$ is 
computationally indistinguishable from the distribution 
defined by $g^x$ with $x \leftarrow  [1..\phi(N)]$. This change does not affect
protocol security arguments. Thus, we do not provide a new proof for APSI in this paper.

P$^3$MT involves two phases: {\em offline} and an {\em online}.

\medskip
\noindent
During the {\em offline} phase:
\begin{enumerate}
\item \CA{} generates RSA public-private keypair $((N,e),d)$,
publishes $(N,e)$, and keeps $d$ private.
\item \Client{} prepares a fingerprint of drug ${\mathcal{D}}$:
$
fp({\mathcal{D}})\equal\{(b^*_j||j) \},
$ 
where each $b^*_j$ is  expected at position $j$ of a genome suitable for ${\mathcal{D}}$.
\item \Client{} obtains from {\sf \small CA} an authorization $auth(fp({\mathcal{D}}))$, where
$
auth(fp({\mathcal{D}})) \equal \{\sigma_j \spa | \spa \sigma_j \equal \mbox{H}(b^*_j||j)^d \bmod N\}. 
$
\item \Server{} runs the offline stage of the APSI protocol in Fig.~\ref{fig:apsi},
on input, $\mathcal{G} \equal \{(b_1||1),\ldots,(b_n||n)\}$, and publishes resulting $\{ts_1,\ldots,ts_n\}$.
\end{enumerate}

\medskip
\noindent
During the {\em online} phase:
\begin{enumerate}
\item \Client{} and \Server{} run the online part of the APSI protocol in Fig.~\ref{fig:apsi}. Recall that \Client's
input is $(fp({\mathcal{D}}), \linebreak auth(fp({\mathcal{D}})))$, and \Server's is $\mathcal{G}$.
\item After the interaction, \Client{} obtains
$fp({\mathcal{D}}) \cap \mathcal{G}$, and uses this information to
determine whether \Server\ is well-suited for drug ${\mathcal{D}}$.
\end{enumerate}

We note that $auth$ is needed to limit the scope of the test on a patient DNA:
the FDA can guarantee that: (i) $fp$ only covers the appropriate set of required nucleotides, 
and (ii) pharmaceuticals cannot input arbitrary portions of a patient genome.

The proposed P$^3$MT protocol is resilient against (randomly distributed) sequencing errors. The
size of the fingerprint input by \Client{} in the protocol is negligible compared to the size of the entire genome.
Thus, positions corresponding to \Client{} input are affected by errors with extremely low probability.

\descr{Performance Evaluation.} To estimate the efficiency of the P$^3$MT protocol, we consider two
genetic tests commonly performed in the context of personalized medicine: the analysis of  
{\em hla-B} and {\em tpmt} genes. Our choice is also motivated by the size of their 
fingerprints that, according to genomics experts, is representative of most personalized medicine tests.

First, we look at the {\em hla-B*5701} allelic variant, one G$\rightarrow$T mutation associated with extreme 
sensitivity to abacavir, a drug  used in HIV treatment~\cite{migueles2000}. In diploid organisms (such as humans),  
mutation may occur in either chromosome inherited from the parents. Thus, the related fingerprint contains 
2 $(${\em nucleotide}, {\em position}$)$ pairs.
We also consider the analysis of {\em tpmt} typically done before prescribing 6-mercaptopurine to 
leukemia patients. As shown in~\cite{yates1997}, two alleles are known to cause the {\em tpmt} 
disorder: (1) one presents a mutation G$\rightarrow$C in position 238 of gene's c-DNA, (2) the other 
presents one mutation G$\rightarrow$A in position 460 and one A$\rightarrow$G in position 
719.\footnote{For more details on {\em tpmt} and c-DNA, refer to~\cite{thiopurine} and~\cite{lewis2003human}, respectively.}
Therefore, the resulting fingerprint contains these 6 $($nucleotide, position$)$ pairs.

In the underlying APSI protocol (Fig.~\ref{fig:apsi}), cryptographic operations on \Server\ genome
do not depend on \Client{} input. Therefore, they can be computed offline, once for all possible tests.
Moreover, we have designed the P$^3$MT protocol to be as generic as possible. Our protocol runs on the whole \Server's genome --- with linear complexity --- in order to address future 
scenarios where genomics advances will cause better understanding of many more regions of human genomes.
To reduce offline costs, we apply reference-based compression~\cite{daily2010data, brandon2009data} -- a technique 
commonly used to efficiently represent genomic information. In particular, \Server{} input consists of all differences between
its genome and the reference sequence.  We emphasize that this technique does not require any biological correctness 
of the reference genome that is only used for compression~\cite{referencegenomecompr}.
This allows us to reduce the size of \Server{} input to about 1\% of the entire genome.

\begin{table}[h]
\centering
\begin{tabular}{|l|l|c|c|c|c|}
\hline
\multirow{2}{*}{\bf Test} &\multirow{2}{*}{\bf Party} & \multicolumn{1}{c|}{\bf Offline} & \multicolumn{2}{c|}{\bf Online} \\
& & {\bf Time} & {\bf Time} & {\bf Size} \\
\hline
\multirow{2}{*}{\emph{hla-b}*5701} & \Client{} & -- & $0.82$ ms & $256$ B\\
& \Server\ & $206$ mins & $0.82$ ms & $4.14$ GB\\
\hline
\multirow{2}{*}{\em tpmt} & \Client{} & -- & $2.46$ ms & $768$ B\\
& \Server\ & $206$ mins & $2.46$ ms & $4.14$ GB\\
\hline
\end{tabular}
\vspace{-0.2cm}
\caption{Computation and communication costs of P$^3$MT protocol for {\em hla-b} (2-nucleotide 
fingerprint) and {\em tpmt} (6-nucleotide fingerprint) tests.}
\label{tab:persmed}
\vspace{-0.0cm}
\end{table}

Table \ref{tab:persmed} summarizes execution time and bandwidth costs
of the P$^3$MT protocol used for testing {\em hla-B} and {\em tpmt}.
These costs cannot be meaningfully 
compared to prior work, since, to the best of our knowledge, 
there is no other technique targeting privacy-preserving personalized medicine testing.
Furthermore, as mentioned in Sec.~\ref{sec:related}, there are no current techniques 
that enforce fingerprint {\em authorization} by a trusted entity, such as 
the FDA. Also, prior work is essentially designed for operation on DNA snippets, and it is unclear
how to efficiently adapt it to full genomes. Although a detailed experimental study is out of
scope of this paper, we intend to include it as part of future work.

\begin{figure*}[t!]
\centering
\fbox{\small
\begin{minipage}{1.75\columnwidth}
\begin{tabular}{lcl}
\textbf{\underline{\Client}}, on input \spa $\Cset=\{c_1,\ldots,c_v\}$ &\hspace{-0.25cm}[Common Input: $(p,q,g,\mbox{H},\mbox{H}')$] &  \textbf{\underline{\Server}}, on input \spa  $\Sset=\{s_1,\ldots, s_w\}$\vspace{0.2cm}\\
&& \underline{\bf Offline} \vspace{0cm}\\
&& $\{\hat s_1,\ldots, \hat s_w\} \leftarrow \Pi (\Sset)$, with $\Pi$\\
&&  random permutation\\
&&$R_s \leftarrow \Z_{q}$\\
&&$\forall j\spa 1\leq j\leq w,\spa {ts_{j}} = \mbox{H}'(\mbox{H}(\hat s_j)^{R_s})$\vspace{-0.5cm}\\
 & \hspace{-0.6cm} $\xymatrix@1@=85pt{& \ar[l]^*
+<4pt>{}_*+<4pt>{\{ts_{1},\ldots,ts_{w} \}}}$\vspace{0.1cm}\\
&\\
\underline{\bf Online} && \\
$\forall i\spa 1\leq i\leq v, R_{c:i} \leftarrow \Z_q$ & \hspace{-0.6cm} $\{a_1,\ldots,a_v\}$ &\underline{\bf Online}\\
$\forall i\spa 1\leq i\leq v, \spa a_i={\mbox{H}(c_i)}^{R_{c:i}}$ &\hspace{-0.6cm}  $\xymatrix@1@=85pt{\ar[r]^*+<4pt>{}&}$ &$\forall i\spa 1\leq i\leq v,\spa a_i' =(a_i)^{R_s} $\vspace{-0.1cm}\\
&\hspace{-0.6cm}  $\xymatrix@1@=85pt{& \ar[l]^*
+<4pt>{}_*+<4pt>{\{a'_{1},\ldots,a'_{v}\}}}$\vspace{-0.3cm}\\
$\forall i\spa 1\leq i\leq v,\spa tc_{i} =\mbox{H}'((a'_i)^{1/R_{c:i}})$\vspace{0.1cm}\\
 {\bf Out:}  $\{c_i | c_i \in \mathcal{C} $ and $tc_i \in \{ts_1,..,ts_w\} \} $&& \\
\end{tabular}
\end{minipage}
}
\caption{PSI Protocol from \cite{stas} (simplified for semi-honest security). It runs on common input of two primes $p$ and 
$q$  (s.t. $q|p-1$), a generator $g$ of a subgroup of size $q$ and two hash functions, H and H$'$, 
modeled as random oracles. All computation is mod $p$.}
\label{fig:psi}
\vspace{-0.2cm}
\end{figure*}
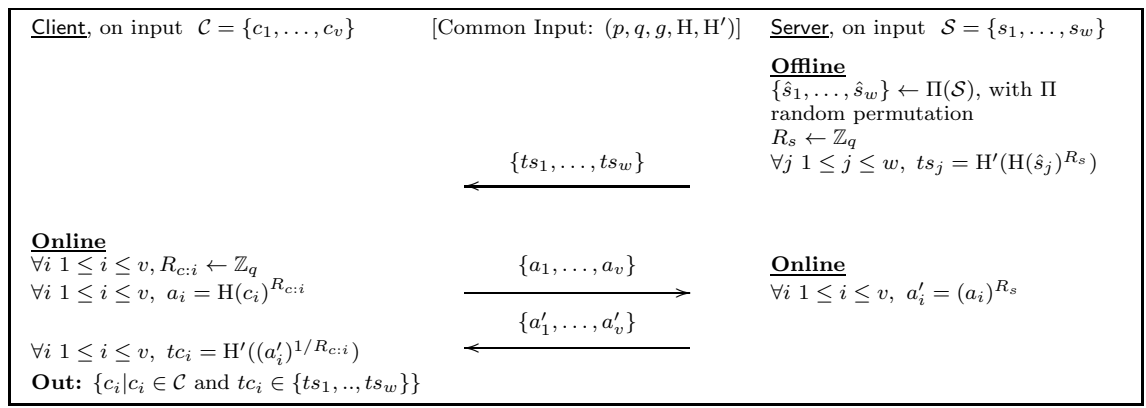

\subsection{Privacy-Preserving Genetic Compatibility Testing}\label{sec:genetic}
Genetic Compatibility Testing (GCT) can predict whether potential partners are at risk of
conceiving a child with a recessive genetic disease. This occurs when both partners carry
at least one gene affected by mutation, i.e., they are either asymptomatic carriers or 
actual disease sufferers. As in the Beta-Thalassemia example discussed in Sec.~\ref{sec:intro},
asymptomatic carriers usually need to learn whether their potential partner is also a carrier
of the same disease, since this would pose a serious risk to their potential off-spring.

To achieve genetic compatibility testing with privacy we introduce the concept of
\textbf{\em Privacy-Preserving Genetic \linebreak Compatibility Testing} (PPGCT) that allows participants to run
GCT without disclosing to each other: (1) any other genomic information,
and (2) which disease(s) they are carrying or  being tested for.

Current biological knowledge of the human genome allows  screening for a
genetic disease associated with one SNP in a specific gene. In other words,
most well-characterized genetic diseases are caused by a mutation in a single gene.
However, we anticipate that, in the near future, researchers will develop tests
for more complex diseases (e.g., diabetes or hypertension) involving multiple genes
and multiple mutations. Therefore, we aim to design PPGCT techniques
not limited to single-mutation diseases.
Additional motivating examples for PPGCT include compatibility testing for sperm and organ
donors.

The proposed PPGCT protocol involves two participants: \Client{} and \Server.
\Client{} runs on input of a fingerprint of a genetic disease $\hat{\mathcal{D}}$.
\Server{} runs on input of its fully-sequenced  genome $\mathcal{G}$.
At the end of the interaction, \Client{} learns the output of the test, i.e.,
whether \Server\ carries disease $\hat{\mathcal{D}}$.

Our cryptographic building block is Private Set Intersection (PSI) \cite{FC10,FNP,stas,AC10}.
We select the specific PSI construction in~\cite{stas}, shown in Fig.~\ref{fig:psi},
since it achieves the best communication and computation complexity.
It can also be instantiated in the malicious model with only a small constant additional overhead.

The PPGCT protocol involves the following steps:
\begin{compactenum}
\item \Client{} builds a fingerprint corresponding to her genetic diseases 
$
fp(\hat{\mathcal{D}})\equal \{(b^*_j||j) \},
$
where each $b^*_j$ {\em  is expected at position $j$ of a genome with disease}
$\hat{\mathcal{D}}$.
\item \Client{} and \Server\ run the PSI protocol in Fig. \ref{fig:psi} on respective inputs:
$fp(\hat{\mathcal{D}})$ and $\mathcal{G}$.
\item \Client{} obtains $fp(\hat{\mathcal{D}}) \cap \mathcal{G}$, and uses this information to
determine whether \Server\ carries disease $\hat{\mathcal{D}}$.
\end{compactenum}

\vspace{0.2cm}
\noindent
The change from PSI-CA to PSI is motivated as follows.
Depending on the disease being tested, a positive outcome occurs if the genome contains either: (1) 
the entire disease fingerprint, or (2) a given subset of nucleotides. 
In case of (1), the test result is positive only if:
$fp(\hat{\mathcal{D}}) \subset \mathcal{G}, \;
\mbox{i.e.,}
\; fp(\hat{\mathcal{D}}) \cap \mathcal{G} \equal fp(\hat{\mathcal{D}})$: 
if this happens, there is actually no difference between the output of PSI and that of PSI-CA.
However, PSI-CA is preferred over PSI since, if the test is negative, less information about
\Server{} genome is revealed to \Client. 
In case of (2), cardinality of set intersection is insufficient to
assess the test result, since \Client{} needs to learn which  fingerprint nucleotides
appear in \Server's genome.

Similar to its P$^3$MT counterpart, the PPGCT protocol is resilient to uniformly distributed 
errors. In particular, since input size of  \Client{} is small, corresponding positions in \Server{} 
genome are affected by errors with very low probability.

\descr{Open Problem:} Unfortunately, a malicious \Client{} could potentially
{\em harvest} \Server's genetic information (in addition to
that needed for the compatibility test) by inflating its input.
For instance, a healthy \Client{} could learn whether or not \Server{} carries a given
genetic disease, unrelated to the compatibility testing. 

\descr{Performance.} As concrete examples, we use genetic compatibility tests for
two genetic disorders: Roberts syndrome and Beta-Thalassemia.
We chose them since they are fairly common and the size of their fingerprints
is representative of that in most genetic compatibility tests.

Similar to P$^3$MT, we stress that cryptographic operations performed 
on \Server{} genome, in the underlying PSI protocol, do not depend on \Client{} input. 
Therefore, these operations can be pre-computed (just once) ahead of time.

First, we consider testing for Roberts syndrome. an autosomal genetic disorder, 
characterized by pre- and post-natal growth deficiency, limb malformations, and distinctive 
skull and facial abnormalities. As shown in~\cite{gordillo2008}, there are 26 single point 
mutations (in the {\em esco2} gene) causing this syndrome. Since humans are diploid organisms, 
we expect Roberts syndrome fingerprint to contain about 52 $($nucleotide, location$)$ pairs.

Next, we turn to Beta-Thalassemia. As pointed out in \cite{HBB}, more than 250 mutations 
in the {\em hbb} gene have been found to cause this disorder and most of them  involve a change 
in a single nucleotide. Although reliable techniques to perform
this test {\em in silico} are not yet available, it is reasonable to assume that the size of the 
Beta-Thalassemia fingerprint would  include 2$\times$250 = 500 $(${\em nucleotide, location}$)$ pairs. 

Table \ref{tab:gentest} summarizes run time (computational) and bandwidth requirements for the PPGCT protocol 
for Roberts syndrome and Beta-Thalassemia, respectively. Following the same arguments as in P$^3$MT experiments,
we let \Server{} input the portion of its genome that differs from the reference genome, i.e., about 1$\%$.

\begin{table}[h]
\centering
\begin{tabular}{|l|l|c|c|c|c|}
\hline
\multirow{2}{*}{\bf Test} &\multirow{2}{*}{\bf Party} & \multicolumn{1}{c|}{\bf Offline} & \multicolumn{2}{c|}{\bf Online} \\
& & {\bf Time} & {\bf Time} & {\bf Size} \\
\hline
\multirow{2}{*}{\em \small Roberts syndrome} & \Client{} & -- & $7.26$ ms & $62.5$ KB\\
& \Server\ & $67$ mins & $7.26$ ms & $4.14$ GB\\
\hline
\multirow{2}{*}{\em \small Beta-Thalassemia} & \Client{} & -- & $70$ ms & $6.5$ KB\\
& \Server\ & $67$ mins & $70$ ms & $4.14$ GB\\
\hline
\end{tabular}
\vspace{-0.2cm}
\caption{Computation and communication costs of the PPGCT protocol for Beta-Thalassemia (500-nucleotide fingerprint)
and Roberts syndrome (52-nucleotide fingerprint) tests.}
\label{tab:gentest}
\vspace{-0.2cm}
\end{table}

Performance of the PPGCT protocol cannot be meaningfully compared to prior work. 
As discussed in Sec.~\ref{sec:related}, it is not trivial to adapt current secure 
pattern matching techniques to genetic compatibility testing on \FSGS.
An experimental study (including the adaptation of such techniques) is left for 
future work.

\begin{figure*}[t!]
\centering
\fbox{\small
\begin{minipage}{1.75\columnwidth}
\begin{tabular}{lcl}
\textbf{\underline{\Client}}, on input \spa $\Cset=\{c_1,\ldots,c_v\}$ &\hspace{-0.25cm}[Common Input: $(p,q,g,\mbox{H},\mbox{H}')$] &  \textbf{\underline{\Server}}, on input \spa  $\Sset=\{s_1,\ldots, s_w\}$\vspace{0.2cm}\\
&& \underline{\bf Offline} \vspace{0cm}\\
&& $\{\hat s_1,\ldots, \hat s_w\} \leftarrow \Pi (\Sset)$, with $\Pi$\\
&&  random permutation\\
&&$R_s \leftarrow \Z_{q}$\\
&&$\forall j\spa 1\leq j\leq w,\spa {ts_{j}} = \mbox{H}'(\mbox{H}(\hat s_j)^{R_s})$\vspace{-0.5cm}\\
 & \hspace{-0.6cm} $\xymatrix@1@=85pt{& \ar[l]^*
+<4pt>{}_*+<4pt>{\{ts_{1},\ldots,ts_{w} \}}}$\vspace{0.1cm}\\
&\\
\underline{\bf Online} && \\
$\forall i\spa 1\leq i\leq v, R_{c:i} \leftarrow \Z_q$ & \hspace{-0.6cm} $\{a_1,\ldots,a_v\}$ &\underline{\bf Online}\\
$\forall i\spa 1\leq i\leq v, \spa a_i={\mbox{H}(c_i)}^{R_{c:i}}$ &\hspace{-0.6cm}  $\xymatrix@1@=85pt{\ar[r]^*+<4pt>{}&}$ &$\forall i\spa 1\leq i\leq v,\spa a_i' =(a_i)^{R_s} $\vspace{-0.1cm}\\
&\hspace{-0.6cm}  $\xymatrix@1@=85pt{& \ar[l]^*
+<4pt>{}_*+<4pt>{\{a'_{1},\ldots,a'_{v}\}}}$\vspace{-0.3cm}\\
$\forall i\spa 1\leq i\leq v,\spa tc_{i} =\mbox{H}'((a'_i)^{1/R_{c:i}})$\vspace{0.1cm}\\
 {\bf Out:}  $\{c_i | c_i \in \mathcal{C} $ and $tc_i \in \{ts_1,..,ts_w\} \} $&& \\
\end{tabular}
\end{minipage}
}
\caption{PSI Protocol from \cite{stas} (simplified for semi-honest security). It runs on common input of two primes $p$ and 
$q$  (s.t. $q|p-1$), a generator $g$ of a subgroup of size $q$ and two hash functions, H and H$'$, 
modeled as random oracles. All computation is mod $p$.}
\label{fig:psi}
\vspace{-0.2cm}
\end{figure*}
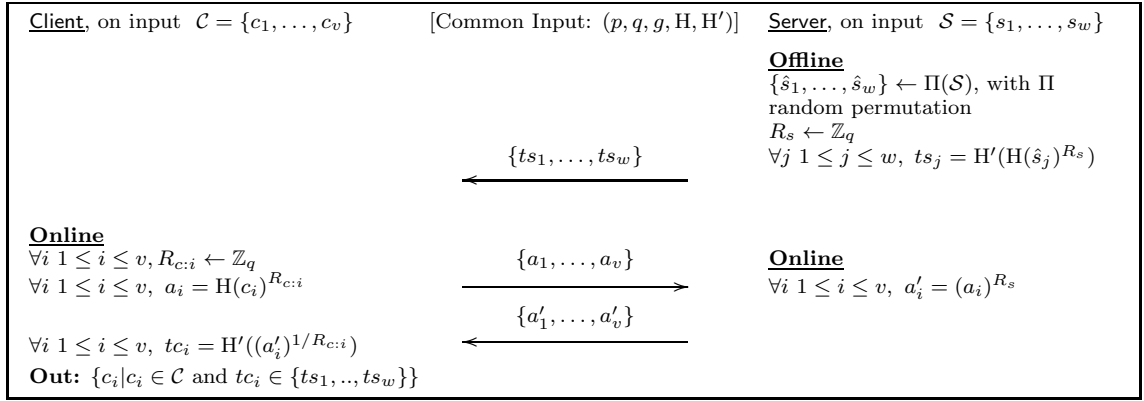

\section{Security Discussion}\label{sec:security}
We now discuss security properties of protocols presented in this paper.
In general, security of each protocol is based on that of the underlying building blocks.
Therefore, we omit proof details to ease presentation.
Also, out cryptographic building blocks  (PSI-CA, APSI, and PSI) can be generally 
used in a {\em black-box} manner. One can select any instantiation without
affecting security of our protocols, as long as the chosen construction yields secure 
PSI/APSI/PSI-CA functionality. However, we pick specific instantiations to maximize
protocol efficiency.
As discussed earlier, we consider semi-honest adversaries (participants). Nevertheless,
we are not restricted to this model, since our cryptographic building blocks are (provably) adaptable
to the malicious participant model, incurring a small constant extra overhead. 

\descr{PPGPT.} We now show that RFLP-based PPGPT protocol (Sec.~\ref{sec:paternity}) is secure 
against semi-honest adversaries. We assume that 
PSI-CA performs secure computation of the $\mathcal{F}_{\sf PSI\mbox{-}CA}$ functionality,
in the presence of semi-honest participants. We select the construction in~\cite{ePrint}, that is secure
under the One-More-DH assumption in the Random Oracle Model (ROM).\\*
We divide the protocol in two phases. In the first, both \Client{} and \Server{} privately and 
independently perform the RFLP-related computation on their
respective inputs. (This covers steps 1 to 3 of PPGPT). At the end of this phase,
\Client{} and \Server{} construct sets $F_C$ and $F_S$, respectively. Clearly, during this phase,
neither participant learns anything about the other's input. During the second phase (steps 4-5),
participants use $F_C$ and $F_S$ as their respective inputs to PSI-CA. Given the security of
the latter, \Client{} only learns $|F_S \cap F_C|$. PSI-CA protocols may reveal $|F_S|$
to \Client{} and $|F_C|$ to \Server{}. However, $|F_S|\equal|F_C| \equal l$, which is already known to both parties.

\descr{P$^3$MT.} Similarly, security of the P$^3$MT protocol (in Sec.~\ref{sec:PM}),
a\-gainst semi-honest  \Client{} and \Server, stems from security of the underlying protocol --- APSI.
That is, if APSI performs secure computation of the $\mathcal{F}_{\sf APSI}$
functionality in the presence of semi-honest participants, then P$^3$MT is also secure.
This holds since a semi-honest participant with a non-negligible advantage in distinguishing between real and 
simulated executions of P$^3$MT would have the same advantage in distinguishing between 
real and simulated executions of APSI.
Although one can use APSI as a black box, for efficiency reasons, we prefer
instantiations that allow pre-computation on \Server{} input. In our instantiation, we select the 
APSI construction in~\cite{AC10}, proven secure under the RSA and DDH assumptions (in ROM).

\descr{PPGCT.} Finally, security of the PPGCT protocol (Sec.~\ref{sec:genetic})
against semi-honest adversaries relies on that of the
underlying PSI protocol, to which it is immediately reducible. (In other words,  a semi-honest participant with a
non-negligible advantage in distinguishing between real and simulated executions of PPGCT would have
the same advantage in distinguishing between real and simulated executions of PSI.)
Again, although one can use PSI as a black box, for efficiency reasons, we need PSI instantiations that 
allow pre-computation on \Server{} input, such as  OPRF-based constructs \cite{HL08,FC10,AC10,stas}.
We chose the PSI  from~\cite{stas}, proven secure under the One-More-DH assumption (in ROM).

\section{Conclusions and Future Work} \label{conc}
This paper identified and explored three popular privacy-sensitive genomic applications: (i) paternity tests,  
(ii) personalized medicine and (iii) genetic compatibility testing. Unlike most previous work, we focused on 
\FSGS. This scenario poses new challenges, both in terms of privacy and 
computational cost. For each application, we proposed an efficient construction, based on well-known 
cryptographic tools: Private Set Intersection (PSI), Private Set Intersection Cardinality (PSI-CA), and 
Authorized Private Set Intersection (APSI). Experiments show that these protocols incur 
online overhead sufficiently low to be practical today.  In particular, our protocol for privacy-preserving 
paternity testing is significantly less expensive --- in both computation and communication --- than prior work. 
Furthermore, all protocols presented in this paper have been carefully constructed to mimic the state-of-the-art
of ({\em in vitro}) biological tests currently performed in hospitals and laboratories.

Items for future work include, but are not limited to\vspace{0.05cm}:
\begin{compactitem}
\item Introducing privacy-preserving genetic paternity testing based on STR and/or SNP comparison.
\item Exploring privacy-preserving techniques to realize genetic ancestry testing, i.e., to discover whether or not
individuals are related up to a certain degree (e.g.,
see 
\url{http://23andme.com}.)
%
%
\item Exploring probabilistic privacy-preserving genetic paternity and ancestry testing based on MinHash techniques,
as discussed in~\cite{espresso}.
\item Extending the paternity test protocol to allow both participants to determine whether the other party introduced
correct input according to some auxiliary authorization. (Note that APSI does not suffice since one of the 
parties might alter its input so that the test is negative). 
\item Investigation of additional privacy-sensitive applications for fully-sequenced genomes, such as 
certified forensic identification, where the subject of investigation must prove the authenticity of its input; privacy-preserving 
organ recipients compatibility, where a subject efficiently identifies a matching sample 
without revealing information about her genome.
\item Extending our experiments to include adaptation of secure pattern matching and text processing to 
personalized medicine and genetic compatibility testing on full genomes.
\end{compactitem}

\descr{Acknowledgements.} 
We are grateful to Christophe Magnan for useful hints about the testing environment and to 
anonymous ACM CCS'11  reviewers for helping us improve the paper.
Work of Pierre Baldi is supported, in part, by grants: NIH LM010235 and NIH-NLM T15 LM07443.

\bibliographystyle{plain}
\bibliography{_bibfile-extended}

\end{document}